\documentclass[prb,aps,floats,amssymb,showkeys,showpacs,preprint,
superscriptaddress,tightenlines]{revtex4}
\usepackage{graphicx}
\usepackage{epsfig,bm}
\begin{document}
\preprint{}

\title{ 
The free energy in a magnetic field and the  universal scaling  \\ 
equation of state for the three-dimensional Ising model\\}

\author{P. Butera\cite{pb}}
\affiliation
{Dipartimento di Fisica Universita' di Milano-Bicocca\\
and\\
Istituto Nazionale di Fisica Nucleare \\
Sezione di Milano-Bicocca\\
 3 Piazza della Scienza, 20126 Milano, Italy}

\author{M. Pernici\cite{mp}} 
\affiliation
{Istituto Nazionale di Fisica Nucleare \\
Sezione di Milano\\
 16 Via Celoria, 20133 Milano, Italy}

\date{\today}
\begin{abstract}
 We have substantially extended the high-temperature and
 low-magnetic-field (and the related low-temperature and
 high-magnetic-field) bivariate expansions of the free energy for the
 conventional three-dimensional Ising model and for a variety of other
 spin systems generally assumed to belong to the same critical
 universality class.  In particular, we have also derived the
 analogous expansions for the Ising models with spin $s=1,3/2,..$ and
 for the lattice euclidean scalar field theory with quartic
 self-interaction, on the simple cubic and the body-centered cubic
 lattices.  Our bivariate high-temperature expansions, which extend
 through $K^{24}$, enable us to compute, through the same order, all
 higher derivatives of the free energy with respect to the field,
 namely all higher susceptibilities.  These data make more accurate
 checks possible, in critical conditions, both of the scaling and the
 universality properties with respect to the lattice and the
 interaction structure and also help to improve an approximate
 parametric representation of the critical equation of state for the
 three-dimensional Ising model universality class.

\end{abstract}
\pacs{  05.50.+q, 64.60.De, 75.10.Hk, 64.70.F-, 64.10.+h}
\keywords{Ising model, high-temperature expansions, magnetic field, 
equation of state}

\maketitle

\small
\section{Introduction}
  We present a brief analysis of high-temperature (HT) and low-field
expansions for the free energy of the conventional 3D Ising model in
an external uniform magnetic field, extended from the presently
available\cite{katsura,mckenzie,essam,esshunt} order $17$ up to order
$24$ in the case of the simple-cubic ($sc$) lattice, and from the
order $13$ up to $24$ in the case of the body-centered-cubic ($bcc$)
lattice.  In addition to the conventional Ising model (i.e. with spin
$s=1/2$), we have considered also a few models with spin $s>1/2$, and
the lattice scalar euclidean field theories with even polynomial
self-interaction.  All results for the simple Ising system in a field
can be readily transcribed into the lattice-gas model language and
therefore are of immediate relevance also for the theory of the
liquid-gas transition\cite{fisherburford,fishertarko}.  The HT and
low-field expansions of the spin-$s$ Ising models can be
transformed\cite{domb_dg,ferer} into low-temperature (LT) and
high-field expansions.

The spin-$s$ Ising model in an external magnetic field $H$ is described by the 
Hamiltonian\cite{wortis,bakerkin,rosksack,nickelrehr}
\begin{equation}
{\cal H}\{s\}=-\frac{J} {s^2}  \sum_{<ij>} s_is_j-\frac{mH}{s}\sum_i s_i
\label{isingesse}
\end{equation}
where $s_i=-s,-s+1,...,s$ is the spin variable at the lattice site $\vec i$,
 $m$ is the magnetic moment of a spin, $J$ is the exchange coupling. 
 The first sum extends over all distinct nearest-neighbor pairs of sites,
 the second sum over all lattice sites. The conventional Ising model 
 is recovered by setting $s=1/2$.

 The one-component self-interacting  lattice scalar field theory is
described by the Hamiltonian\cite{luscher,campo2002,bcfi4}
\begin{equation}
{\cal H}\{\phi\}= -\sum_{<ij>} \phi_i \phi_j+\sum_i (V(\phi_i) + H\phi_i).
\label{pdifi}
\end{equation}
Here $-\infty <\phi_i< +\infty$ is a continuous variable associated to
 the site $\vec i$ and $V(\phi_i)$ is an even polynomial in the
 variable $\phi_i$. In this study, we have only considered the
 specific model in which $ V(\phi_i)= \phi^2_i + g(\phi^2_i-1)^2$, although
 we can  cover interactions of a more general form.

All these models are expected to belong to the $3D$ Ising universality
 class, therefore our eextensive set of series expansion data can be
 used to test the accuracy of the basic hypotheses of critical scaling
 and universality with respect to the lattice and the interaction
 structure, by comparing the estimates of the exponents and of
 universal combinations of critical amplitudes for the various models
 as well as by forming approximate representations of the equation of
 state (ES).  In this study our attitude\cite{bcfi4,bcspinesse} is, to
 some extent, complementary to the current one. Usually, universality
 is essentially assumed from the outset: for example, in the
 renormalization group (RG)
 approach\cite{brez,zinnbook,gz1,gz2,zinnrep,soko,denjoe,morris,tetra},
 an appropriate scalar field theory in continuum space is taken as the
 representative of the Ising universality class, as suggested by the
 independence of the renormalization procedure from the details of the
 microscopic interaction.  Also in HT and MonteCarlo approaches,
 attention has been recently
 focused\cite{campo2002,bcfi4,bcspinesse,blt,engels} on particular
 continuous- or discrete-spin lattice models which exhibit vanishing
 (or very small) leading non-analytic corrections to
 scaling\cite{zinn,chenfisher} in order to be able to estimate more
 accurately the physical quantities of interest.  In this report, we
 prefer to take advantage of our extended expansions to test a wide
 sample of models, expected to belong to the same universality class,
 and to show how closely, already at the present orders of expansion,
 each model approaches the predicted asymptotic scaling and
 universality properties.

The paper is organized as follows: in Sect.II we briefly characterize
our expansions, sketch the method of derivation and list the numerous
tests of correctness passed by the series coefficients.  In Sect. III
we define the higher-order susceptibilities, whose critical parameters
enter into the determination of the scaling equation of state and
update an approximate representation of it. In Sect. IV we discuss
numerical estimates of exponents, amplitudes and universal
combinations of these, that can be computed from the bivariate
series. In the last section, we summarize our results and draw some
conclusions.
\section{ Extensions of the bivariate series expansions}
 The HT series expansion coefficients for the models under study have
 been derived by a fully computerized algorithm based on the
 vertex-renormalized linked-cluster (LC) method which calculates the
 mean magnetization per spin in a non-zero magnetic field from the set
 of all topologically distinct, connected, 1-vertex-irreducible (1VI),
 single-rooted graphs\cite{wortis}.  We have taken advantage of the
 bipartite structure of the $sc $ and the $bcc$ lattices to restrict
 the generation of graphs to the subset of the bipartite graphs,
 i.e. to the graphs containing no loops of odd length.

In the past, the LC method was employed mainly to
derive expansions in the absence of magnetic field.
  In the presence of a field, the most 
extensive\cite{katsura,mckenzie} data so far available in $3D$ were
derived indirectly, by transforming\cite{domb_dg,ferer}
  bivariate LT and high-field expansions\cite{sykes} 
 into HT and low-field
expansions.  This computation was  performed only for the
$s=1/2$ Ising model, although some LT and high-field data existed also for
other values of the spin.  Shorter HT  expansions in a finite field 
had also been previously obtained\cite{fishertarko}, only for the
$s=1/2$ model, by a direct expansion of the free energy. It is worth
noting that we are now in a position to follow the opposite route:
namely of transforming  our bivariate HT data for the spin-$s$
Ising systems into LT and high-field expansions, thus 
extending the known  results.

Our improvements of the presently available HT series in a field are
 summarized in Table \ref{tab1}, in the case of the $sc$ and the $bcc$
 lattices.  Similar extensions for the same class of models, in the
 case of the simple quadratic ($sq$) lattice and for bipartite
 lattices in $d>3$ space dimensions, will be discussed elsewhere.  The
 series expansions coefficients will be tabulated in a separate paper.

 The feasible correctness checks of our computations are inevitably
 partial, since the extended expansions include information much wider
 than that already available in the literature.  The easiest
 non-trivial check is that our procedure yields the known bivariate
 expansion of the free energy for the spin $1/2$ Ising model in a
 finite field on the one-dimensional lattice. We have also checked
 that our results agree, through their common extent, with the old
 data cited above\cite{katsura,mckenzie} for the spin $1/2$ Ising
 system in a magnetic field, both on the $sc$ and the $bcc$ lattices.
 Otherwise, our results can only be compared with the related data in
 zero field, in particular with the HT expansions of the free energy
 and its second field-derivative, both for the Ising model with
 general spin-$s$ and for the scalar field model, on the $sc$ and the
 $bcc$ lattices, which have been tabulated\cite{bcfi4,bcspinesse}
 through order $K^{25}$, while the 4th field-derivative is already
 known\cite{bcfi4,bcspinesse} through $K^{23}$ for both lattices.  Our
 results agree, through their common extent, also with the expansions
 of the 6th field derivative in zero field, tabulated\cite{camposerie}
 up to order $K^{19}$, and of the 8th field-derivative,
 tabulated\cite{camposerie} up to order $K^{17}$, in the case of the
 spin $1/2$ model on the $bcc$ lattice. We have finally checked that
 our expansions reproduce the $sc$ lattice calculations of the 6th
 field-derivative (known up to order $K^{19}$ ), of the 8th (known up
 to order $K^{17}$) and of the 10th (known up to order $K^{15}$) in
 the case of the $sc$ lattice scalar field with quartic coupling
 $g=1.1$, which have been tabulated in Ref.[\onlinecite{campo2002}].

\begin{table}
\caption{ Maximal order in $K$ of the high-temperature and low-field
 expansions of the free energy for the models in the Ising
 universality class considered in this note.  }
\begin{tabular}{|c|c|c|}
 \hline
           & Existing data[\onlinecite{mckenzie}]  & This work\\
 \hline
$sc$ lattice& &\\
 \hline 
Ising $s=1/2$&        17 &     24\\
Ising $s>1/2$&       0  &     24\\
 $\phi^4 $ & 0 &     24\\
 \hline
$bcc$ lattice& &\\
 \hline 

Ising $S=1/2$ &   13       &      24\\ 
Ising  $s>1/2$&       0  &     24\\
 $\phi^4 $ & 0 &     24\\
     
 \hline
\colrule  
\end{tabular} 
\label{tab1}
\end{table}

It is fair to remark that the finite-lattice (and the related
 transfer-matrix) methods of expansion\cite{enting,arisue} have shown
 more efficient\cite{baxterent} than the LC approach, at least for
 $s=1/2$ in $d=2$ dimensions, even in presence of a magnetic field,
 while they remain rather difficult and unpractical in higher space
 dimensions. In the case of the two-dimensional spin $1/2$ Ising
 model, with the support of a variational approximation, these methods
 made a representation of the ES of unprecedented accuracy\cite{baz}
 possible. In the future, these techniques might prove to be
 competitive\cite{arisue} in the $3D$ case also for calculations in a
 nonvanishing field. We believe, however, that it has been worthwhile
 to test and develop also a LC approach, which expresses the
 series coefficients in terms of polynomials in the moments of the
 single-spin measure and therefore, unlike the finite lattice method,
 is flexible enough to apply also to non-discrete state models such as
 the one-component scalar field model\cite{bcfi4} within the Ising
 universality class studied here and, more generally, to the
 O(N)-symmetric spin\cite{bc8212} or lattice-field systems in any
 space dimension.

\subsection{ The algorithms}

To give an idea of the strong points of our graphical algorithms, we
mention that, using only an ordinary desktop personal 
quad-processor-computer with a $4G$ fast memory (RAM), our code can complete in
seconds all calculations already documented\cite{katsura,mckenzie} in
the literature (see Table \ref{tab1}).  The whole renormalized
calculation presented here, can be completed in a CPU-time of a few
days, most of which goes in producing the highest order of expansion.
In what follows all timings are single-core times.

\begin{table}
\caption{ 
The number of simple, connected,  bipartite, unrooted 1VI graphs 
with $l$ lines
 and with a given number $v$  of odd vertices, 
  which contribute to the  HT expansion coefficient of the free energy
 at order $K^{l}$ }
\center
\begin{tabular}{|c|c|c|c|c|c|c|c|c|c|c|}
 \hline
&   $v$ &&&&&&&&& \\     
 \hline
$l$ &  0&  2&  4 &6&8&10&12&14&16& totals\\
 \hline
4& 1& 0& 0& 0& 0& 0& 0& 0& 0& 1\\
5& 0& 0& 0& 0& 0& 0& 0& 0& 0& 0\\
6& 1& 1& 0& 0& 0& 0& 0& 0& 0& 2\\
7& 0& 1& 0& 0& 0& 0& 0& 0& 0& 1\\
8& 2& 1& 1& 0& 0& 0& 0& 0& 0& 4\\
9& 0& 3& 1& 1& 0& 0& 0& 0& 0& 5\\
10& 3& 6& 5& 0& 0& 0& 0& 0& 0& 14\\
11& 0& 11& 7& 2& 0& 0& 0& 0& 0& 20\\
12& 9& 20& 31& 4& 1& 0& 0& 0& 0& 65\\
13& 0& 49& 53& 22& 0& 0& 0& 0& 0& 124\\
14& 20& 101& 194& 54& 7& 0& 0& 0& 0& 376\\
15& 0& 258& 432& 238& 20& 2& 0& 0& 0& 950\\
16& 84& 520& 1471& 732& 127& 0& 0& 0& 0& 2934\\
17& 0& 1482& 3725& 2886& 434& 29& 0& 0& 0& 8556\\
18& 300& 3243& 12233& 9531& 2403& 97& 5& 0& 0& 27812\\
19& 0& 9646& 33608& 36067& 9675& 845& 0& 0& 0& 89841\\
20& 1520& 21859& 109796& 123543& 46241& 4023& 133& 0& 0& 307115\\
21& 0& 68697& 318283& 460225& 191416& 26435& 594& 13& 0& 1065663\\
22& 8186& 163780& 1048349& 1608030& 858792& 134409& 6672& 0& 0& 3828218\\
23& 0& 533569& 3166399& 5970246& 3566324& 757696& 40686& 744& 0&
14035664\\
24& 52729& 1328836& 10594514& 21241772& 15475018& 3796365& 317259& 4267&
38& 52810798\\
 \hline
\colrule  
\end{tabular} 
\label{tab2}
\end{table}

The LC computation has been split into three parts.  First, we
generated the simple, bipartite, unrooted, topologically distinct 1VI
graphs.  This part is memory intensive, but takes only a few
hours. Table \ref{tab2} lists the numbers of these graphs from order 4
through 24. In a second step, we computed the single-rooted
multigraphs, their symmetry numbers and their lattice embeddings. This
part of the calculation requires little memory: in the case of the
$bcc$ lattice, completing the 24th order took approximately a day,
while in the case of the $sc$ lattice two weeks were necessary. In the
latter case, most of the time was spent to determine the graph
embeddings.  These two parts of the calculation were implemented by
C++ codes, and used the ``Nauty'' \cite{nauty} library to compute the
graph certificates and symmetry factors. The relevant procedures of
this package were supplemented with the GNU Multiprecision Arithmetics
Library\cite{gnu} to get the exact graph symmetry numbers.  The third
step implements the algebraic vertex-renormalization\cite{wortis}
procedure by deriving the magnetization from the single-rooted 1VI
graphs and then, by integration, the free energy ${\cal F}(K,h)=\sum
f_n(h)K^n$. Here $K=J/k_BT$, with $k_B$ the Boltzmann constant and $T$
the temperature, while $h=mH/k_BT$ is the reduced magnetic field.  The
magnetization is expressed in term of the bare vertices $M^0_{i}(h)$
obtained deriving $i$ times with respect to $h$ the generating
function $M^0_{0}(h)= {\rm ln}\Big[\frac{{\rm sinh}(h(2s+1)/2s)} {{\rm
sinh}(h/2s)} \Big]$ in the case of the spin-$s$ Ising systems, or, in
the case of the scalar field system, $M^0_{0}(h)= {\rm ln}\Big[ \int
d\phi e^{-V(\phi)+h\phi} \Big]$.  For example, the HT expansion
coefficient of the free energy at order $K^2$, on the $bcc$ lattice,
is given by
\begin{equation}
f_2(h) =  2(M^0_{2})^{2}(h) +32(M^0_{1})^{2}(h)M^0_{2}(h)
\end{equation}
while on the  $sc$ lattice
\begin{equation}
f_2(h) = \frac{3} {2}(M^0_{2})^{2}(h) +18(M^0_1)^2(h)M^0_2(h).
\end{equation}
Table \ref{tab3} lists the number of monomials of the bare vertices,
with a given number $v$ of odd indices which contribute to the free
energy HT expansion coefficient at order $K^{l}$. Equivalently, this
is the number of admissible vertex-degree sequences of the (far more
numerous) graphs contributing to this coefficient.  Notice that the
monomials containing at least two bare vertices of odd order are the
overwhelming majority. They all vanish in zero field, which shows that
the finite field calculation has a substantially higher complexity. The 
renormalization through 24th order was performed in a few hours.  The
third step of the calculation is based on codes written in the Python
and Sage\cite{sage} languages.

\begin{table}
\caption{ The number  of monomials in the bare vertices, 
 with a given number $v$  of odd vertices, 
  which contribute to the  HT expansion coefficient of the free energy
  on a bipartite lattice, at order $K^{l}$ }
\center
\begin{tabular}{|c|c|c|c|c|c|}
 \hline
&   $v$ &&&& \\     
 \hline
$l$ &  0&  2&  4 &  $v>4$& totals\\
 \hline
1&  0& 1& 0&  0&1\\
2&  1& 1& 0&  0&2\\
3&  0& 3& 1&  0&4\\
4&  3& 4& 3 & 0&10\\
5&  0& 10& 6 &2&18\\
6 & 6& 14& 15& 6&41\\
7&  0& 27& 25& 18&70\\
8&  14& 39& 45& 39&137\\
9&  0& 70& 77& 86&233\\
10& 25& 94& 130& 164&413\\
11& 0& 157& 201& 305&663\\
12& 53& 222& 318& 541&1134\\
13& 0& 348& 481& 924&1753\\
14& 89& 457& 742& 1529&2817\\
15& 0& 699& 1091& 2519&4309\\
16& 167& 941& 1589& 3972& 6669\\
17& 0 &1379& 2289& 6213& 9881\\
18& 278& 1796& 3314& 9566& 14954\\
19& 0& 2577& 4635& 14487& 21699\\
20& 480& 3370& 6492& 21662& 32004\\
21& 0& 4711& 9010& 32134& 45855\\
22& 760& 5965& 12430& 46887& 66042\\
23& 0& 8257& 16858& 67949& 93064\\
24& 1273& 10664& 22895& 97543& 132375\\
 \hline
\colrule  
\end{tabular} 
\label{tab3}
\end{table}

It is also not without interest that in a preliminary step of our
 work, we have been able to employ the simple {\it unrenormalized}
 linked-cluster method\cite{wortis}, which uses {\it all}
 topologically distinct unrooted connected graphs (including
 multigraphs) to compute the bivariate expansions of the free energy
 through order 20. It takes only one day to complete this calculation.
 Of course, while the unrenormalized procedure is algebraically
 straightforward, it would make further extensions of the series
 unpractical, using our desktop computers, for the rapid increase with
 order of the combinatorial complexity and, as a consequence, of the
 memory requirements. The computation of the 21st order does not fit
 in 4 GB of RAM, but would require some increase of memory.  These
 calculations are however interesting by themselves, both because the
 unrenormalized method is still generally (and too pessimistically)
 dismissed as unwieldy beyond just the first few orders, and because
 they provide a valuable cross check, through order 20, of the results
 of the algebraically more complex vertex-renormalized procedure,
 which remains necessary to push the calculation to higher orders.

\section{ Asymptotic scaling  and the equation of state}
The hypothesis of asymptotic
scaling\cite{widom,dombhunter,pata,kada,fish67} for the singular part
${\cal F}_s (\tau,h)$ of the reduced specific free energy, valid as both $h$
and $\tau$ approach zero, can be expressed in the form
\begin{equation}
{\cal F}_s (\tau,h) \approx |\tau|^{2-\alpha} Y_{\pm}(h/|\tau|^{\beta \delta}).
\label{freescaling}
\end{equation}
where $\tau=(1-T_c/T)$ is the reduced temperature.  The exponent
$\alpha$ specifies the divergence of the specific heat, $\beta$
describes the small $\tau$ asymptotic behavior of the spontaneous
specific magnetization $M$ on the phase boundary $(h \rightarrow 0^+,
\tau<0)$
\begin{equation}
M \approx B(-\tau)^{\beta}
\label{ampmagsp}
\end{equation}
 with  $B$  the critical amplitude of $M$. 
 The exponent 
 $\delta$ characterizes the small $h$ asymptotic behavior of the
magnetization on the critical isotherm $(h \ne 0,\tau=0)$, 
\begin{equation}
|M| \approx B_c |h|^{1/\delta}
\label{amphiso}
\end{equation}
 and $B_c$ is the corresponding critical amplitude.
For the exponents $\alpha$ and $\beta$,
 we have assumed the values $\alpha=0.110(1)$ and
 $\beta=0.3263(4)$, obtained using the scaling and hyperscaling 
relations, from the 
   HT estimates\cite{bcfi4}  of the susceptibility exponent 
$\gamma=1.2373(2)$ and of the correlation-length exponent $\nu=0.6301(2)$.

 The functions $Y_{\pm}(w)$ are defined for $0\le w \le \infty$ and
have a power-law asymptotic behavior as $w \rightarrow \infty$.  The
$+$ and $-$ subscripts indicate that different functional forms are
expected to occur for $\tau<0$ and $\tau>0$.  The usual scaling laws
follow from eq.(\ref{freescaling}). The simplest consequence of
eq.(\ref{freescaling}), which will be tested using our HT expansions,
is that the critical exponents of the successive derivatives of ${\cal
F}_s (\tau,h)$ with respect to $h$ at zero field, are evenly spaced by
the quantity $\Delta=\beta \delta$, usually called ``gap exponent''.
More precisely, let us define the zero-field $n$-spin connected
correlation functions at zero wavenumber (also called {\it higher
susceptibilities} when $n>2$) by the equation
\begin{equation}
 \chi_{n}(K)=(\partial^n {\cal F}(h,K)/\partial h^n)_{h=0}
=\sum_{s_2,s_3,...,s_{n}}<s_1 s_2...s_{n}>_c.
\label{ncorr}
\end{equation}
For odd values of $n$, these quantities vanish in the symmetric HT
phase, while they are nontrivial for all $n$ in the broken-symmetry LT
phase.  For even values of $n$ in the symmetric phase, and for all $n$
in the broken phase, scaling implies that, as $T\rightarrow T_c^+$
along the critical isochore ($h=0, \tau>0$) or, as $T\rightarrow
T_c^-$ along the phase boundary, we have
\begin{equation}
 \chi_{n}(\tau) \approx C^{\pm}_{n}|\tau|^{-\gamma_{n}}(1 +
 b^{\pm}_{n} |\tau|^{\theta} + \ldots)
\label{2ncorras}
\end{equation}
 where $\gamma_{n}=\gamma +(n-2)\Delta$, $b^{\pm}_n$ and $\theta$ are,
respectively, the amplitude and the exponent which characterize the
leading non-analytic correction to asymptotic scaling.  The
value\cite{deng} $\theta= 0.52(2)$ has been estimated for the
universality class of the $3D$ Ising model.  Assuming also the
validity of hyperscaling, we can conclude that $2\Delta=3\nu+\gamma$.

An important bonus of our bivariate calculations, is the significant
 extension the HT expansions of the higher susceptibilities. We have
 added one more term to the existing\cite{bcspinesse} HT expansion of
 $\chi_{4}(K)$, five terms to that\cite{camposerie} of $ \chi_{6}(K)$,
 seven to that\cite{campo2002} of $\chi_{8}(K)$ and nine to that of $
 \chi_{10}(K)$.  In the case of the susceptibilities of order $2n>10$,
 no data at all were available so far.  We have now extended,
 uniformly in the order, the HT expansions of all higher
 susceptibilities $ \chi_{2n}(K)$ with $2n \ge 4$.  In this paper, we
 shall present only a preliminary analysis of these quantities, while
 a more detailed discussion of our bivariate expansions will be
 postponed to a forthcoming article.

The scaling form of the equation of state $M= {\cal M}(h,T)$, relating
 the external reduced magnetic field $h$, the reduced temperature $\tau$ and
 the magnetization $M$, when $h$ and $\tau$ approach zero, is simply
 obtained by differentiating eq.(\ref{freescaling}) for 
$f_s(\tau,h)$ with respect to $h$ 
\begin{equation}
  M\approx -|\tau|^{\beta}Y^{(1)}_{\pm}(h/|\tau|^{\beta \delta})
\label{eqstatwidom}
\end{equation}
Here we have used the relation $\gamma=\beta(\delta -1)$.  By further
differentiation of eq. (\ref{eqstatwidom}) with respect to the field,
also the higher susceptibilities are recognized to have a scaling form
\begin{equation}
\chi_n(h,\tau)= (\partial^{n-1} M/\partial h^{n-1}) \approx  
-|\tau|^{-\gamma_n} Y^{(n)}_{\pm}(h/|\tau|^{\beta \delta}).
\label{chiscaling}
\end{equation}
 The hypothesis of universality states that, in addition to the
critical exponents, the function $Y_{\pm}(w)$, (and therefore also
 its $n-th$ derivative  $Y^{(n)}_{\pm}(w)$) is
universal\cite{watson} up to multiplicative constants ({\it metric
factors\cite{ahapriv}}) which fix the scales of $h$ and $\tau$ in each
particular model within a universality class.  Accordingly, one can
conclude that a variety of dimensionless combinations of critical
amplitudes are universal.

The ES can also be written in the equivalent
 form\cite{dombhunter,widom,griffiths}
\begin{equation}
 h(M,\tau) \approx M|M|^{\delta-1}f(\tau/|M|^{1/\beta})
\label{eqstatgrif}
\end{equation}
in which a single scaling function $f(x)$, universal up to metric
 factors, describes both the regions $\tau<0$ and $\tau>0$.  The
 function $ h(M,\tau)$ is known\cite{griffiths} to be regular analytic
 in a neighborhood of the critical isotherm and of the critical
 isochore.  From general thermodynamic arguments\cite{griffiths} one
 can infer that $f(x)$ is a positive monotonically increasing regular
 function of its argument, in some interval $-x_0 \leq x \leq \infty$,
 with $x_0 >0$. Moreover $f(-x_0)=0$.
  The local behavior of the function $f(x)$ can be further 
 determined, by the requirement of consistency with the scaling laws,
 in terms of critical amplitudes of quantities computable from our HT
 and LT series.  By differentiating this form of the ES with respect
 to $M$, we get the asymptotic behavior $f(x) \propto x^{\gamma}$, for
 large positive $x$.  Setting $\tau=0$, the ES reduces to
 eq.(\ref{amphiso}) and $f(0)=B_c^{-\delta}$. If $h \rightarrow 0$ at
 fixed $\tau<0$, we expect to find a nonvanishing spontaneous
 magnetization $M$, therefore the ES implies that $f(x)$ must
 vanish. Since $f(-x_0)=0$, we have $-\tau/M^{1/\beta}= x_0$ and, from
 eq.(\ref{ampmagsp}), we conclude that $x_0=B^{-1/\beta}$.  We can then
 fix the metric factors by normalizing the field to $B_c^{-\delta}$
 and the reduced temperature to $B^{-1/\beta}$.  The 
 expansion of $f(x)$ for large positive
 $x$ is expressed in terms of the critical parameters
 characterizing the HT side of the critical point.  The small $x$
 expansion, which uses the parameters of the critical isotherm, and
 the negative $x$ region related to the parameters of the LT side of
 $T_c$, will be discussed in a forthcoming paper presenting our
 analysis of the extended LT expansions.

Summarizing the more detailed discussion of
Ref.[\onlinecite{gz1}], we can also observe that, in the large positive $x$
(small magnetization) region, where the magnetic field $ h(M,\tau) $
has a convergent expansion in odd powers of $M$, the ES is more
conveniently expressed in terms of the variable 
$z= M\tau^{-\beta} x_0^{\beta}$.
The ES takes then the form
\begin{equation}
 h(M,\tau) = \bar h|\tau|^{\beta\delta}F(z)
\label{eqstatgriz}
\end{equation}
where $\bar h$ is a constant and $F(z)$ is normalized by the equation 
$F'(0)=1$.
The small $z$ expansion of $F(z)$ can be written as

\begin{equation}
 F(z)=z+\frac{1}{6}z^5+F_5z^5+F_7z^7+...
\label{fz}
\end{equation}

The coefficients $F_5,F_7,...$ are defined by the equation
$F_{2n-1}=r^+_{2n}/(2n-1)!$, in terms of the ratios $r^+_{2n}$ which
will be introduced in the next section. They have been computed within
the RG approach\cite{gz2,soko}, by the $\epsilon$-expansion
($\epsilon=4-d$) up to five loops, by the perturbative $g$-expansion
at fixed dimension $d=3$ up to the same order, by other RG
approximations\cite{denjoe,morris,tetra}, by HT
expansions\cite{bcg2n,campo2002}, by MonteCarlo
methods\cite{tsyp,kim}.  Our estimates of the first few $r^+_{2n}$ by
extended HT expansions, are reported in Table \ref{tab9}.
\subsection{ A parametric form of the ES}
A parametric form\cite{scho1,scho2,jos}
 has been introduced to formulate an approximate
representation of the ES in the whole critical region and as an aid in
the comparison with the experimental data.  The parametrization is
chosen to embody the analyticity properties of $h(M,\tau)$ and the
scaling laws. These properties make the parametric form convenient to
approximate the ES in the whole critical region by using only  
 an HT input, such as the small $z$ expansion eq.(\ref{fz}) of $F(z)$.
In this approach, the scaled field and the reduced temperature are
 expressed as the following functions
 
\begin{equation}
M=m_0 R^{\beta}\theta
\label{param}
\end{equation}

\begin{equation}
\tau=R(1-\theta^2)
\label{paratau}
\end{equation}

\begin{equation}
h=h_0R^{\beta\delta}l(\theta)
\label{parah}
\end{equation}

of generalized radial and angular coordinates
 $R \ge 0$ and $ -\theta_0 \le \theta \le \theta_0 $, with $\theta_0>
 1$  the smallest positive zero of the function $l(\theta)$.
  The radial coordinate $R$ measures the distance in the $h,T$
 plane from the critical point, and the angular coordinate $\theta$
 specifies a direction in this plane.  Therefore, $\theta=0$
 corresponds to the critical isochore, $\theta=\pm 1$ is associated to
 the critical isotherm and $\theta=\pm \theta_0$ to the coexistence
 curve.  
The function $l(\theta)$, normalized by $l'(0)=1$, is
   odd  and regular   for $|\theta| < \theta_0$, as implied by the 
regularity of $f(x)$ and the invertibility of the above variable
 transformation in this interval.

The variable 
 $z$ is  then expressed as
\begin{equation}
z=\frac{\rho \theta}{(1-\theta^2)^{\beta}}
\label{zeta}
\end{equation}
and the function $F(z)$ of eq.(\ref{fz}) is related to $l(\theta)$ by

\begin{equation}
l(\theta)=\frac{1}{\rho}((1-\theta^2)^{\beta+\gamma}F(z(\theta))
\label{ltheta}
\end{equation}

Here $\rho=m_0x_0^{\beta}$ is a positive constant related to the
 arbitrary normalization constant $m_0$ appearing in
 eq. (\ref{param}).  If $F(z)$ were exactly known, the corresponding
 $l(\theta)$ given by eq. (\ref{ltheta}) should not depend on $\rho$.
 However the polynomial truncations of $l(\theta)$ that can be formed
 from the first few available terms of the expansion eq. (\ref{fz}) of
 $F(z)$ will have coefficients $l_{2n+1}(\rho)$ depending not only on
 the coefficients $F_5, F_7,...$ and on the exponents $\beta$ and
 $\gamma$, but also on $\rho$.

In particular\cite{gz1},  expanding both 
sides of eq.(\ref{ltheta}), one obtains:
\begin{equation}
l_3(\rho)=\frac{1}{6}\rho^2 - \gamma
\label{l3}
\end{equation}
\begin{equation}
l_5(\rho)=\frac{1}{2}\gamma(\gamma-1)+\frac{1}{6}(2\beta-\gamma)\rho^2
+F_5\rho^4
\label{l5}
\end{equation}

\begin{equation}
l_7(\rho)=-\frac{1}{6}\gamma(\gamma-1)(\gamma-2)+\frac{1}{12}
(2\beta-\gamma)(2\beta-\gamma+1)\rho^2+ (4\beta-\gamma)F_5\rho^4 +F_7\rho^6
\label{l7}
\end{equation}

 \begin{eqnarray}
 l_9(\rho)=\frac{1}{24}\gamma(\gamma-1)(\gamma-2)(\gamma-3)+\frac{1}{36}
 (2\beta-\gamma)(2\beta-\gamma+1)(2\beta-\gamma+2)\rho^2+\\
 \label{l8}
\nonumber
  \frac{1}{2}(4\beta-\gamma)(4\beta-\gamma+1)F_5\rho^4
  +(6\beta-\gamma)F_7\rho^6 +F_9\rho^8
 \end{eqnarray}

  etc.  

\begin{figure}[tbp]
\begin{center}
\leavevmode
\includegraphics[width=3.37 in]{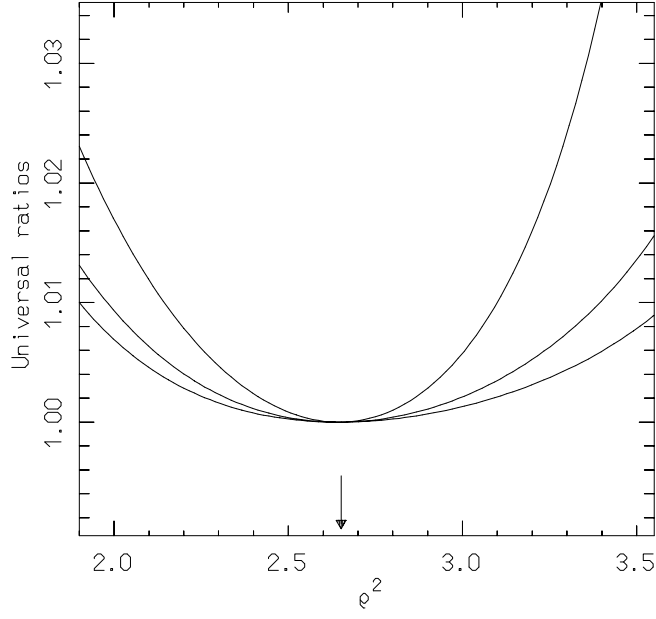}
\caption{\label{rapp} A plot vs the parameter $\rho^2$, of the
 universal combinations of critical amplitudes $C^+_2/C^-_2$(upper
 curve), $C^+_2B^{\delta-1}/ B_c^{\delta}$(middle curve) and
 $C^+_4B^2/(C^+_2)^3$(lower curve) obtained from the truncated
 polynomial approximation of $l(\theta)$ eq. (\ref{applth}). The
 computation is based on the coefficients $F_{2n-1}$ with $n=1,..7$,
 estimated in this work.  For convenience, the curves are normalized to
 their minimum values. }
\end{center}
\end{figure}

The dependence on $\rho$ of the coefficients $l_{2n+1}$ has been
 exploited to improve the approximation of $l(\theta)$. A first
 approach consists in fixing $\rho$ to the value $\rho_m$ which
 minimizes\cite{gz1} the modulus of the highest-order expansion
 coefficient $l_{2n+1}(\rho)$ of $l(\theta)$ that can be determined
 reliably from the available coefficients $F_{2n-1}$.  A second
 method\cite{campo2002} is based on computing some universal
 combinations of critical amplitudes in terms of $l(\theta)$ and then
 in choosing for $\rho$ the unique value that makes all such
 quantities stationary. We may follow this route and consider, for
 example, the dependence on $\rho$ of the universal ratio of the
 susceptibility amplitudes above and beneath $T_c$, namely
 $C^+_2/C^-_2$, and of the ratios $C^+_4B^2/(C^+_2)^3$ and
 $C^+_2B^{\delta-1}/ B_c^{\delta}$.  If we plot these quantities vs
 $\rho^2$, we obtain Fig.\ref{rapp}, which  indicates that the
 choice $\rho^2=2.615$ should be optimal. Then, using the central
 values both of the coefficients $F_{2n-1}$ up to $n=7$, as obtained from
 our Table \ref{tab9}, and of the exponents $\beta$ and $\gamma$ 
as indicated above and  fixing $\rho$ to its optimal value, the
 following  form of $l(\theta)$ can be determined
 \begin{equation}
 l(\theta) \approx \theta -0.8014(50)\theta^3+ 0.00946(30)\theta^5+ 
0.00141(40)\theta^7+0.00029(10)\theta^9-0.00011(5)\theta^{13}
 \label{applth}
 \end{equation}
Here we have neglected the term in $\theta^{11}$, whose coefficient is
$O(10^{-6})$, and have indicated the last three terms only to show
that their contribution in the interval of interest $|\theta| <
\theta_0$ is very small.  The function $l(\theta)$ vanishes at
$\theta=\theta_0 \approx \pm 1.1273$.

The analogous result for this auxiliary function obtained in
Ref.[\onlinecite{gz2}], fixing $\rho$ by
 the first method and choosing the values
 $\beta=0.3258(14)$ and $\gamma=1.2396(13)$ of the critical exponents, is
 \begin{equation}
 l(\theta)\approx  \theta -0.762(3)\theta^3+ 0.0082(10)\theta^5 
 \label{applthzinn}
 \end{equation}
 which vanishes at $\theta\approx \pm 1.1537$. In this case, 
the coefficients $F_{2n-1}$
 were obtained by a RG five-loop perturbation expansion in $d=3$. On
 the other hand, computing the $F_{2n-1}$ by the RG
 $\epsilon$-expansion to fifth order 
and choosing $\beta=0.3257(25)$ and
 $\gamma=1.2355(50)$ leads\cite{gz2}  to:
 \begin{equation}
 l(\theta)\approx  \theta -0.72(6)\theta^3+ 0.0136(20)\theta^5 
 \label{applthzinn2}
 \end{equation}
 More recently 
in Ref.[\onlinecite{campo2002}],
 using values of the exponents very near to those used in our paper,
 and deriving the $F_{2n-1}$  from an HT
 expansion of $sc$-lattice scalar-field models with self-couplings
 appropriately chosen to suppress the leading correction to scaling,
 the following expression was obtained, 
 \begin{equation}
 l(\theta) \approx 
\theta-  0.736743\theta^3+ 0.008904\theta^5 - 0.000472\theta^7
 \label{applthpi}
 \end{equation}
 which vanishes at $\theta\approx \pm 1.1741$. As stressed in
 Refs. [\onlinecite{gz1,gz2}], the 
 alternative forms eqs. (\ref{applth}), (\ref{applthzinn}),
 (\ref{applthzinn2}) and (\ref{applthpi}) cannot be directly compared, 
 because they are associated to different parametrizations (different
 values of $\rho$).  One should rather compare the universal
 predictions obtained from them, for example, for the universal
 amplitude combinations, which we have reported in Table
 \ref{tab4}. In this Table, $A^+$ and $A^-$ denote the amplitudes of
 the specific heat above and beneath $T_c$.  Our estimates are
 compared with the corresponding ones obtained\cite{gz2} from the
 polynomials $l(\theta)$ in eq.(\ref{applthzinn}) and
 eq.(\ref{applthzinn2}) and with those obtained\cite{campo2002} from
 eq. (\ref{applthpi}).  The results from the various methods show a
 good overall consistency.  In Table \ref{tab4}, we have not reported
 MonteCarlo estimates, which also are
 available\cite{engels,feng,campb} for the ratios $A^+/A^-$,
 $C_2^+B^{\delta-1}/B_c^{\delta}$ and $C_2^+/C_2^-$.  It is worth to
 remark that the computation of the first quantity is difficult,
 because the weak singularity of the specific heat forces to extend
 the simulation very close to the critical point.  Recent
 estimates\cite{feng,campb} of this ratio, in the range
 $0.532(7)-0.540(4)$, based on simulations of large lattices, might
 now supersede older results, which were $\approx 6\%$ larger, thus
 improving the agreement with the ES estimates of Table \ref{tab4}.
 Also the second ratio, involving the amplitude $B_c$ of
 eq.(\ref{amphiso}) is difficult to compute by simulations, for
 similar reasons.  The result
 $C_2^+B^{\delta-1}/B_c^{\delta}=1.723(13)$ of
 Ref.[\onlinecite{engels}], is somewhat larger than the estimates from
 the ES.  In the case of the third ratio, recent
 simulations\cite{campb,engels} have changed the previous larger
 estimates to values in the range $4.67(3)-4.72(11)$, closer to the ES
 results. In the Table \ref{tab4}, we also have not shown the few
 available experimental estimates of some of these combinations, nor
 those, tabulated in Refs.[\onlinecite{gz1,gz2,campo2002}], which are
 based on direct evaluations of the amplitudes by LT and HT
 expansions.  They are completely compatible with the ES results of
 Table \ref{tab4}, but the comparison is not stringent, due to the
 large uncertainties. Therefore we plan to improve the series
 determinations of the amplitudes on the critical isotherm and on the
 coexistence curve, exploiting our extended bivariate LT expansions of
 the free energy for the spin-$s$ Ising models.  A more detailed
 analysis of our results, along with estimates of other universal
 amplitude combinations, and a wider comparison among the results in
 the literature is deferred to a forthcoming paper.

\begin{table}
\caption{ A few universal amplitude combinations  obtained in this
work from the parametric form eq.(\ref{applth}) of the ES. For comparison, we
have  reported also the results obtained from: the slightly different
parametric form eq. (\ref{applthpi}) 
of the ES in Ref.[\onlinecite{campo2002}] based on shorter
 HT expansions and from the parametric forms  
 in Ref.[\onlinecite{gz2}] obtained, either
  from the RG $\epsilon-$expansion or
from the $g-$expansion and using values of the exponents slightly different
 from those used to get the estimates in the first two columns.}
\center
\begin{tabular}{|c|c|c|c|c|}
 \hline
Universal ratios &This work
&Ref.[\onlinecite{campo2002}] &$\epsilon-$expans.[\onlinecite{gz2}] 
&$g-$expans.[\onlinecite{gz2}] \\
 \hline
$A^+/A^-$&0.530(3) &0.529(6)&0.527(37)&0.537(19)\\
$C_2^+/C_2^-$&4.78(3)&4.78(5)&4.73(16)&4.79(10)\\
$C_4^+/C_4^-$ &-9.2(3)&-9.3(5)&-8.6(1.5) &-9.1(6) \\
$-C_4^+B^2/(C_2^+)^3$ &7.8(1)&7.83(4)&8.24(34) & \\
$C_2^+B^{\delta-1}/B_c^{\delta}$ &1.66(2)&1.665(10)&1.648(36)&1.669(18) \\
$\alpha A^+C_2^+/B^2$&0.0563(5)&0.0562(1)&0.0569(35)&0.0574(20) \\
$-C_3^-B/(C_2^-)^2$&6.015(15)&6.018(20)&6.07(19)&6.08(6) \\
 \hline
\colrule  
\end{tabular} 
\label{tab4}
\end{table}

\subsection{ The HT zero-momentum renormalized couplings}
The HT expansions of the higher susceptibilities\cite{dombhunter} will
 be used to evaluate their critical amplitudes $C^{+}_n$ and
 correspondingly the critical limits of the zero-momentum $n-$spin
 dimensionless HT renormalized couplings (RCC's).  These quantities
 enter into the approximate forms of the scaling ES
 eqs. (\ref{eqstatwidom}) and (\ref{eqstatgrif}).

In the HT phase  the first few  $2n$-spins RCC's
are defined as the critical limits as $K \rightarrow K_c^-$ 
of the following expressions
\begin{equation}
 g^+_4(K)=-\frac{V} {\xi^3(K) } \frac{\chi_4(K)}
 { \chi_2^2(K)}
\label{g4}
\end{equation}
\begin{equation}
 g^+_6(K)=\frac{V^2} {\xi^6(K)}\Big [ -\frac{\chi_6(K)} {\chi_2^3(K)}
 + 10 \Big (\frac{\chi_4(K)} { \chi_2^2(K)}\Big)^2 \Big]
\label{g6}
\end{equation}
\begin{equation}
 g^+_8(K)=\frac{V^3} {\xi^9(K)}\Big [ -\frac{\chi_8(K)} 
{\chi_2^4(K)}
+56 \frac{\chi_6(K) \chi_4(K)} {\chi_2^5(K)}
 -280 \Big (\frac{\chi_4(K)} {\chi_2^2(K)}\Big)^3 \Big]
\label{g8}
\end{equation}

\begin{eqnarray}
 g^+_{10}(K)=\frac{V^4} {\xi^{12}(K)}
\Big [-\frac{\chi_{10}(K)} {\chi_2^5(K)}+
 120\frac{\chi_8(K) \chi_4(K)} {\chi_2^6(K)}
+126 \frac{\chi_6^2(K)} { \chi_2^6(K)}\\
\nonumber
 -4620 \frac{\chi_6(K) \chi_4^2(K)} { \chi_2^7(K)}
 +15400 \Big (\frac{\chi_4(K)} {\chi_2^2(K)}\Big)^4 \Big] 
\label{g10}
\end{eqnarray}

\begin{eqnarray}
 g^+_{12}(K)=\frac{V^5} {\xi^{15}(K)}
\Big [-\frac{\chi_{12}(K)} {\chi_2^6(K)}+
 220\frac{\chi_{10}(K) \chi_4(K)} {\chi_2^7(K)}
+792 \frac{\chi_8(K)\chi_6(K)} { \chi_2^7(K)}\\
\nonumber
-17160 \frac{\chi_8(K)\chi_4^2(K)} { \chi_2^8(K)}
-36036 \frac{\chi_6^2(K) \chi_4(K)} { \chi_2^8(K)}
 +560560 \frac{\chi_6(K)\chi_4^3(K)} {\chi_2^9(K)}
-1401400\Big (\frac{\chi_4(K)} {\chi_2^2(K)}\Big)^5 \Big]
\label{g121}
\end{eqnarray}

\begin{eqnarray}
 g^+_{14}(K)=\frac{V^6} {\xi^{18}(K)}
\Big [-\frac{\chi_{14}(K)} {\chi_2^7(K)}
+364\frac{\chi_{12}(K) \chi_4(K)} {\chi_2^8(K)}
-50050\frac{\chi_{10}(K) \chi_4^2(K)} {\chi_2^9(K)}\\
\nonumber
+2002 \frac{\chi_{10}(K) \chi_6(K)}{\chi_2^8(K)} 
+1716\frac{\chi_8^2(K)} { \chi_2^8(K)}
+3203200\frac{\chi_8(K)\chi_4^3(K)} { \chi_2^{10}(K)}
-360360\frac{\chi_8(K)\chi_6(K)\chi_4(k)}{\chi_2^9(K)}\\
\nonumber
-126126\frac{\chi_6^3(K)}{\chi_2^9(K)}
+10090080 \frac{\chi_6^2(K)\chi_4^2(K)}{\chi_2^{10}(K)}
-95295200 \frac{\chi_6(K)\chi_4^4(K)}{\chi_2^{11}(K)}
+190590400 \Big (\frac{\chi_4}{\chi_2^2(K)} \Big)^6 \Big]
\label{g14}
\end{eqnarray}

Here $\xi(K)$ is the second moment correlation-length, defined by
\begin{equation}
 \xi^2= \frac{\mu_2} {6\chi_2}
\label{xi2}
\end{equation}
 with $\mu_2$ the second moment of the correlation function expressed as
\begin{equation}
 \mu_2= \sum_{s_x} x^2 <s_0 s_x>_c.
\label{mu2}
\end{equation}
Both the HT expansions of $\chi(K)$ and $\mu_2(K)$ 
are tabulated\cite{bcspinesse} through order $K^{25}$
 for the spin-$s$ Ising models.

 The volume $V$ per lattice site takes the value 1 for the $sc$
 lattice and $4/3\sqrt3$ for the $bcc$ lattice. The
 definitions of the quantities $ g^+_{2n}(K)$ given here differ by a
 factor $(2n)!$ from those of Ref.[\onlinecite{bcg2n}]. 

Also the quantities,
\begin{equation}
 I_{2n+4}(K)=\frac{ \chi^n_2(K) \chi_{2n+4}(K)}{\chi^{n+1}_4(K)} 
\label{watso}
\end{equation}
 with $n \geq 1 $, whose critical values are the universal amplitude
 combinations first described\cite{watson} in the literature, and the
 closely related quantities
\begin{equation}
r_{2n}(K)=\frac {g_{2n}(K)}{g_{4}(K)^{n-1}}
\label{r2n}
\end{equation}
which share the computational advantage of being independent of the
 correlation length, will be of relevance in what follows.  The finite 
 critical limits  $g^{+}_{n}$, $r^+_{2n}$ and $I^+_{2n+4}$ of the 
 RCC's, of the ratios $r^+_{2n}(K)$ and of the quantities $I_{2n+4}(K)$, 
 represent universal combinations of HT amplitudes that should be considered 
together with those listed in Table  \ref{tab4}.
  We have not included the expressions of
 higher-order RCC's, because, in spite of our extensions, the
 available series might not yet be long enough to determine safely
 their critical limits.  One should notice that, from the point of
 view of numerical approximation, the $g_{2n}^{+}$, and also the
 quantities derived from them like $r^{+}_{2n}$, are difficult to
 compute, unless $2n$ is small, because they result from relatively
 small differences between large numbers.  These estimates can be
 reliable provided that the uncertainties of the large numbers are
 much smaller than their difference.  For the same reason, these
 quantities are notoriously even more difficult to compute by
 stochastic methods.
\section{ Methods and results of the series analysis}
\subsection{Extrapolation methods}
In the numerical analysis of the series expansions of physical 
  quantities, we shall follow two  procedures aimed to
 determine the critical parameters, namely the values of these
 quantities at the critical point, whenever they are finite, or if
 they are singular, the locations, amplitudes and exponents of the
 critical singularities on (or nearby) the convergence disk in the
 complex $K$ plane.

 A first procedure used in our series analysis, is the differential
 approximant (DA) method\cite{gutt}, a generalization of the well
 known Pad\'e approximant method\cite{gutt}, having a wider range
 of application. In this approach, the values of the quantities or the
 parameters of the singularities can be estimated from the solution,
 called differential approximant, of an initial value problem for an
 appropriate ordinary linear (first- or higher-order) inhomogeneous
 differential equation. This equation has polynomial coefficients
 defined in such a way that the series expansion coefficients of its
 solution equal, up to a certain order, those of the series under
 study.  The various possible equations, and therefore the various DAs
 that can be formed by this prescription, are usually identified by
 the sequence of the degrees of the polynomial coefficients of the
 equation. The approximants are called first-order, second-order DAs
 etc., according to the order of the defining equation.  The
 convergence of the procedure in the case of the Ising models can be
 improved by first performing in the series expansions the variable
 transformation\cite{rosk}
\begin{equation}
z=1-(1-K/K_c)^{\theta}
\label{rosk}
\end{equation}
aimed at reducing the influence of the leading corrections to scaling.
 Here $\theta $ is the exponent which characterizes these corrections.
A sample of estimates of the parameters of the critical singularity is
obtained from the computation of many ``quasi diagonal'' DAs, namely
approximants with small differences among the degrees of the
polynomial coefficients of the defining differential equation, which
use all or most of the given series coefficients.  A first estimate of
a parameter, along with its uncertainty, results from computing the
sample average and standard deviation.  The result can then be
improved by discarding from the sample single estimates which appear
to be obvious outliers, and recomputing the average of the reduced
sample.  A conventional guess of the uncertainty of the parameter
estimate is finally obtained simply, and rather roughly, as a small
multiple of the spread of the reduced sample around its mean
value. This  subjective prescription might, to some extent, allow for
the difficulty to infer possible systematic errors, and to extrapolate
reliably a possible residual dependence of the estimate on the maximum
order of the available series.

A second approach is based on a faster converging modification of the
 standard analysis of ratio-sequence of the series coefficients and
 will be denoted here as the {\it modified ratio approximant}
 (MRA)\cite{zinn,gutt} technique.  Let us assume that the singularity
 of the series expansion of a physical quantity, which is nearest to
 the origin of the complex $K$ plane, is the critical singularity,
 located at $K_c$ and characterized by the critical exponent $\lambda$ 
and the exponent 
$\theta$ of the leading correction to scaling, (this hypothesis is
 generally not satisfied for the LT series). Then eq.(\ref{2ncorras})
 implies the following large $r$ behavior of the series coefficients
 $c_r$
\begin{equation}
c_r= C \frac{r^{\lambda-1}}{\Gamma(\lambda)}K_c^{-r}\Big[1+
\frac{\Gamma(\lambda)}{\Gamma(\lambda-\theta)}
\frac{b} {r^{\theta}}+O(1/r)\Big ]
\label{asycoeff}
\end{equation}
 In this case, the MRA method  evaluates 
$K_c$ by estimating the large $r$ limit of 
the approximant sequence 
\begin{equation}
\big(K_c\big)_r=\Big(\frac{c_{r-2}c_{r-3}}{c_rc_{r-1}} \Big)^{1/4}
exp\Big[\frac{s_r+s_{r-2}}{2s_r(s_r-s_{r-2})}\Big]
\label{mra}
\end{equation}
with
\begin{equation}
s_r=\Big[ {\rm ln}\Big( \frac{c^2_{r-2}}{c_rc_{r-4}}\Big)^{-1}+
{\rm ln}\Big( \frac{c^2_{r-3}}{c_{r-1}c_{r-5}}\Big)^{-1} \Big]/2.
\label{sn}
\end{equation}
By using the asymptotic form Eq.(\ref{asycoeff}), we can
obtain\cite{bcspinesse}  the large $r$ 
asymptotic behavior of the sequence of MRA
 approximants of the critical inverse temperature 
\begin{equation}
\big(K_c\big)_r=K_c \Big[1 -\frac{\Gamma(\lambda)}
{2\Gamma(\lambda-\theta)}\frac{\theta^2(1-\theta)b}{r^{1+\theta}}+
O(1/r^2)\Big]
\label{Knasy}
\end{equation}

The method  estimates also the critical exponent $\lambda$ 
from the sequence
\begin{equation}
 (\lambda)_r= 1+ 2\frac{(s_r+s_{r-2})} {(s_r-s_{r-2})^2}
\label{mraesp}
\end{equation}
with $s_r$ defined by eq.(\ref{sn}).
 In this case the large $r$ asymptotic behavior
 of the sequence $(\lambda)_r$ is
\begin{equation}
 (\lambda)_r=\lambda -\frac{\Gamma(\lambda)}
{\Gamma(\lambda-\theta)}\frac{\theta(1-\theta^2)b}{r^{\theta}}+
O(1/r)
\label{espasy}
\end{equation}
 If the available series expansions are sufficiently long (how long
cannot unfortunately be decided {\it a priori}), the estimates of the
critical points and exponents obtained from extrapolations based on
eqs. (\ref{Knasy}) and (\ref{espasy}) can be competitive in precision
with those from DAs.  If, on the other hand, the series are only
moderately long or the exponent $\lambda>>1$, then corrections of
order higher than $1/r^{1+\theta}$ in eq.(\ref{Knasy}) (or higher than
$1/r^{\theta}$ in eq.(\ref{espasy})) might still be non-negligible.
The same remark applies if the $O(1/r)$ terms in eq.(\ref{asycoeff})
are not sufficiently small.  Therefore, in some cases,
eqs.(\ref{Knasy}) and (\ref{espasy}) might be inadequate to
extrapolate the behavior of the few highest-order terms of the MRA
sequences.
\subsection{Critical parameters of the higher susceptibilities}
For both methods sketched in the previous paragraph, the main
 difficulties of the numerical analysis of the HT expansions are
 related to the presence of the leading non-analytic corrections to
 scaling which appear in the near-critical asymptotic forms of all
 physical quantities. It was however observed\cite{zinn,chenfisher}
 that the amplitudes of these corrections are non-universal and
 therefore, by studying families of models expected to belong to the
 same universality class, one might be able to single out special
 models for which these amplitudes have a very small or vanishing
 size. These models would then be good candidates for a high-accuracy
 determination of the critical parameters of interest. In the
 literature, various models which share this property to a good
 approximation, have been subjected to analysis: among them, the
 lattice $\phi^4$ model on the $sc$ lattice with the value $g=1.1$ of
 the quartic self-coupling\cite{campo2002,hasenb}, or the same model
 on the $bcc$ lattice with the coupling\cite{bcfi4} $g=1.85$. Also the
 spin-$s=1$ and $s=3/2$ Ising systems on the $bcc$
 lattice\cite{bcspinesse}, show very small corrections to scaling. All
 these models will be considered here.

 An accurate estimate $2\Delta = 3.1276(8)$ of the gap exponent which
 improves the four-decade-old\cite{esshunt} estimate $2\Delta =
 3.126(6)$, based on 12th order series, had been already obtained from
 the known 23rd order HT expansions of $\chi_4(K)$ for the spin-$s$
 Ising models\cite{bcspinesse} and for the lattice scalar
 field\cite{bcfi4}, on the $sc$ and $bcc$ lattices. The addition of a
 single coefficient to the expansion of $\chi_4(K)$ does not urge
 resuming a full discussion of the estimates of this exponent and of
 the validity of hyperscaling on the HT side of the critical point,
 already tested with good precision in
 Refs.[\onlinecite{bcfi4,bcgren,bcspinesse}].

\begin{figure}[tbp]
\begin{center}
\leavevmode
\includegraphics[width=3.37 in]{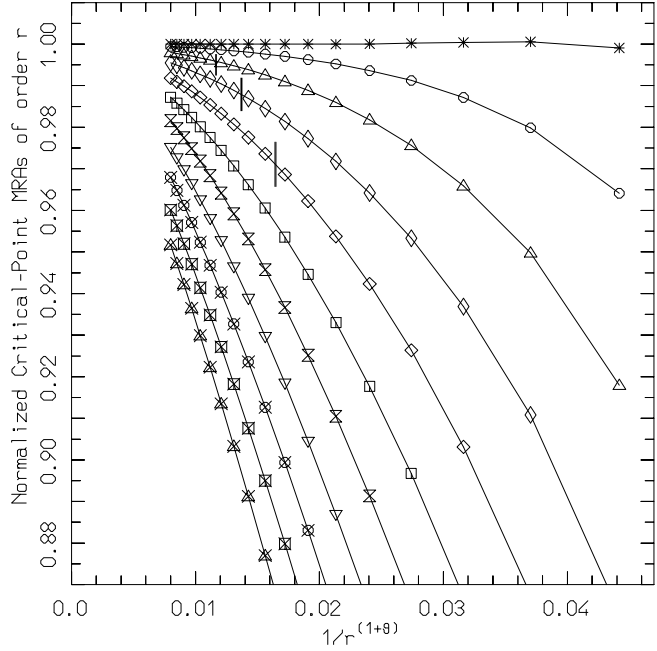}
\caption{\label{chi_2n_fi4_sc_Kc} The sequences of modified ratio
 approximants (MRAs) of the critical point for the scalar field model
 with $g=1.1$ on the $sc$ lattice, plotted vs $1/r^{1+\theta}$. Here
 $r$ is the order of the approximant and $\theta$ is the exponent of
 the leading correction to scaling. We have normalized the MRAs to the 
 estimated value of $K_c$.  The MRAs are obtained from the HT
 expansions of $\chi_2(K)$ (stars), $\chi_4(K)$ (circles), $\chi_6(K)$
 (triangles), $\chi_8(K)$ (rhombs), $\chi_{10}(K)$ (rotated squares),
 $\chi_{12}(K)$ (squares), $\chi_{14}(K)$ (double triangles),
 $\chi_{16}(K)$ (rotated triangles), $\chi_{18}(K)$ (crossed circles),
 $\chi_{20}(K)$ (crossed squares), $\chi_{22}(K)$ (crossed triangles).
 The symbols representing the MRAs are connected by straight lines as
 an aid to the eye.  Small vertical segments on the third, fourth and
 fifth curve from above, indicate the order at which our extension of
 the   $\chi_6(K)$, $\chi_8(K)$, and $\chi_{10}(K)$ 
series begins to contribute to the MRAs.  The six lowest curves
 refer to higher susceptibilities for which  no data exist in the
 literature.}
\end{center}
\end{figure}

 To get some feeling of the reliability of the estimates that can be
 obtained from a study of our HT expansions of the higher
 susceptibilities $\chi_{2n}(K)$, it is convenient to test how
 accurately the critical inverse temperature $K_c$ and the critical
 exponents $\gamma_{2n}$ can be determined from them by using MRAs.
 Let us for example consider the above mentioned self-interacting
 lattice scalar field model of eq.(\ref{pdifi}) on the $sc$ lattice
 with quartic coupling $g=1.1$.  In Fig. \ref{chi_2n_fi4_sc_Kc}, we
 have plotted vs $r^{1+\theta}$ the sequences of the MRA estimates
 $(K_c)_r$ for $K_c$, as obtained from the HT expansions of
 $\chi_{2n}(K)$ with $2n=2,4,...22$.  The MRA sequences are normalized
 by the appropriate limiting value of the sequence $(K_c)_r$,
 estimated in Ref.[\onlinecite{bcspinesse}] and reported in Table
 \ref{tab5}, to make them easily comparable with the corresponding
 sequences obtained from other models in the same universality class.
 The choice of the plotting variable is suggested by
 eq. (\ref{Knasy}).  For the susceptibilities of order $2n \gtrsim 6$
 the curves indicate the presence of strong corrections
 $O(r^{-\sigma})$, with $\sigma$ between 3 and 5, and show that simply
 using eq.(\ref{Knasy}), at the present orders of expansion, would be
 inadequate for extrapolating to $r \rightarrow \infty $ the MRA
 sequences.  No significant quantitative difference in behavior is
 observed in the analogous plots for the other models examined in this
 study, even for those with non-negligible amplitudes of the leading
 corrections to scaling.  On the contrary, in other cases, for example
 for the Ising model with spin-$s=1/2$ or $s=1$ on the $bcc$ lattice,
 the convergence looks even slightly faster. From these plots one may
 conclude that, as the order $2n$of the susceptibility $\chi_{2n}(K)$
 grows, increasingly long expansions are needed\cite{bcg2n} in order
 that the MRA sequences reach the asymptotic form eq. (\ref{Knasy})
 and therefore a given precision can be achieved in the estimate of
 $K_c$.  The general features of this behavior can be tentatively
 explained arguing\cite{bcg2n} that the dominant contributions to the
 HT expansion of $\chi_{2n}(K)$, at a given order $K^r$, come from
 those spin correlation functions in the sum of eq.(\ref{ncorr}), for
 which the average distance among the spins is $\approx r/2n$.
 Accordingly, it seems that the presently available expansions of the
 quantities $\chi_{2n}(K)$, in spite of having the same number of
 coefficients, might not have the same ``effective length'', because
 they describe systems which are, in some sense, rather ``small'', the
 more so the larger is $2n$ . One might conclude that the estimates of
 the critical quantities, derived from the $\chi_{2n}(K)$, should
 probably be taken with some caution for large $n$, even in the case
 of models with very small leading corrections to scaling. However, in
 what follows, we shall observe that,  in some cases, 
in spite of these difficulties,
 the DAs seem to yield smooth and reasonable
 extrapolations of these series to the critical point.

\begin{table}
\caption{Estimates[\onlinecite{bcfi4,bcspinesse}] of the critical
inverse-temperatures $K_c$ used in our study of the Ising systems with
spin $s$  and of the lattice scalar field systems, 
on the $sc$ and the $bcc$ lattices.}
  \center
\begin{tabular}{|c|c|c|c|c|c|c|c|}
 \hline
 &  $s=1/2 $ &$s=1$ &  $s=3/2 $&$ s=2$ &  $s=5/2 $&$s=3$&$\phi^4$   \\
 \hline
$K^{sc}_c$ &0.221655(2)&0.312867(2)&0.368657(2)&0.406352(3)&0.433532(3)&
0.454060(3)&0.375097(1)\\
$K^{bcc}_c$ &0.1573725(10)&0224656(1)&0.265641(1)&0.293255(2)&0.313130(2)
&0.328119(2)&0.2441357(5)\\
\colrule  
\end{tabular} 
\label{tab5}
\end{table}

\subsection{ Scaling and the gap exponent}
\begin{figure}[tbp]
\begin{center}
\leavevmode
\includegraphics[width=3.37 in]{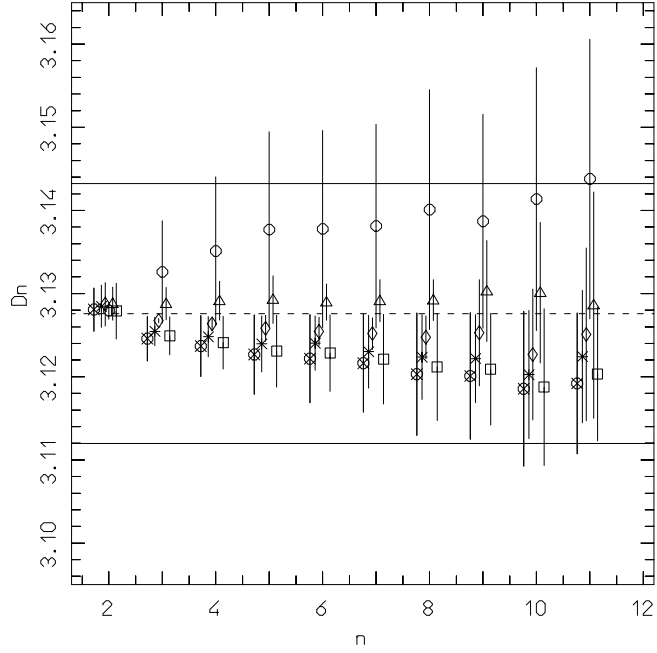}
\caption{\label{diffga} As a simple consequence of the scaling
hypothesis, the exponent differences $D_n=\gamma_{2n}-\gamma_{2n-2}$
should not depend on $n$ and equal $2\Delta$. Here they are obtained
by forming second-order DAs of the ratios
$\chi_{2n}(K)/\chi_{2n-2}(K)$ for $n=2,3,4,...,11$, obtained from the
expansions of the $bcc$ lattice Ising model with spin $s=1/2$
(circles), $s=1$ (triangles), $s=3/2$ (rhombs), $s=2$ (stars), $s=5/2$
(squares) and $s=3$ (crossed circles).  For each value of $n$, the
various symbols have been slightly shifted apart only to avoid
cluttering and keep the uncertainty of each estimate visible.  The
dashed horizontal line represents the estimated\cite{bcspinesse} value
$2\Delta=3.1276(8)$ of twice the gap exponent.  The continuous
horizontal lines indicate a deviation of $0.5\%$ from the expected
central value.}
\end{center}
\end{figure}

\begin{figure}[tbp]
\begin{center}
\leavevmode
\includegraphics[width=3.37 in]{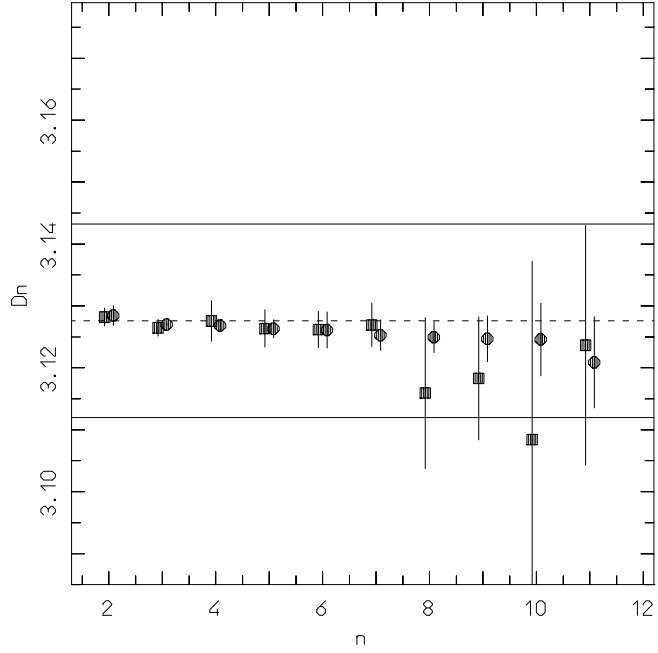}
\caption{\label{diffgafi4} Same as Fig.\ref{diffga}. In this case, the
 differences $D_n=\gamma_{2n}-\gamma_{2n-2}$ have been computed from
 the HT expansions of the higher susceptibilities for the lattice
 scalar field theory with values of the quartic coupling, $g=1.1$ on
 the $sc$ lattice (black squares) and $g=1.85$ on the $bcc$ lattice
 (black circles). These values of $g$ are chosen to minimize the
 leading corrections to scaling.  As in Fig.\ref{diffga}, for each
 value of $n$, the symbols have been slightly shifted apart.  The
 dashed horizontal line represents the estimated\cite{bcspinesse} value
 $2\Delta=3.1276(8)$ of twice the gap exponent.  The continuous
 horizontal lines indicate a deviation of $0.5\%$ from the central
 value.}
\end{center}
\end{figure}

For Ising models on the bcc lattice with spin $s=1/2,1,...,3$, we have
computed the sequences of estimates of the exponent differences
$D_n=\gamma_{2n}-\gamma_{2n-2}$, with $n=2,3,..11$.  These estimates
are obtained from second-order DAs of the ratios
$\chi_{2n}(K)/\chi_{2n-2}(K)$, which use at least 19 series
coefficients. We have imposed that the critical inverse temperatures,
for the various spin systems, take the appropriate
values\cite{bcspinesse}, listed in Table \ref{tab5}.  In
Fig.\ref{diffga}, the exponent differences $D_n$ are plotted vs $n$.
We have observed above that, as a simple consequence of the scaling
hypothesis, when the maximum order of the available HT series grows
large, the $D_n$ should all converge to the same value, equal to twice
the gap exponent $\Delta$, thus being independent of the order $2n$ of
the higher susceptibilities entering into the calculation.  For some
particular values of the spin, e.g. $s=1$ and $s=3/2$, our central
estimates depart by less than $0.1\%$ from the expected result, for
all values of $n$ considered here.  For other values of the spin,
e.g. $s=1/2$, a residual spread of the data remains, which however is
quite compatible with the errors due to the finite length of the
series and to the likely presence of sizable corrections to scaling,
particularly when $n$ is large.  This computation can be repeated,
with similar results, but somewhat larger error bars, for the spin-$s$
Ising system on the $sc$ lattice.  We have shown in
Fig.\ref{diffgafi4} the results of the same computation for the two
lattice scalar field systems with suppressed leading corrections to
scaling, studied here.  In view of the above remarks concerning the
effective length of the expansions of the $\chi_{2n}(K)$, our results
confirm the expectation that, in general, the uncertainties of the
results should grow with $n$.  It should be noted that, both in the
case of the spin-$s$ Ising system and of the scalar field system, the
$bcc$ lattice expansions have a distinctly smoother and more
convergent behavior than for the $sc$ lattice, on a wider range of
values of the order $2n$ of the susceptibilities, probably because the
coordination number of the $bcc$ lattice is larger.  In conclusion,
our results support the validity of the scaling property, while the
rather accurate independence of the estimates of the gap exponent on
the lattice structure and, in the case of the Ising models, on the
value $s$ of the spin, is a valuable indication of universality.
Finally, it is worth to stress that the results of Fig. \ref{diffga}
and Fig. \ref{diffgafi4} for $n \gtrsim 4$ would be difficult to
obtain by numerical approaches other than series expansions.
\subsection{The ratios $r^+_{2n}$ and the critical amplitudes of 
the higher susceptibilities} In Ref.[\onlinecite{esshunt}] the
critical amplitudes $C^{+}_{2n}$ of the higher susceptibilities were
estimated from 12th order series, for the simple $s=1/2$ Ising model,
assuming the now outdated values $\gamma=5/4$ and $\Delta=25/16$ for
the exponents. These series were not long enough that any estimate of
the uncertainties could be tried.  It is then worthwhile to update the
estimates of these amplitudes by using our longer expansions and
biasing the extrapolations by the more precise modern estimates of the
exponents and the critical temperatures cited above.  We can moreover
obtain the corresponding informations also for the other models under
scrutiny.  The critical amplitudes $C^{+}_{2n}$ with $n>2$ can also be
evaluated, with results consistent within their errors , in terms of
$C^{+}_2$, $C^{+}_4$ and of the universal critical values $I^+_{2n+4}
$ of the quantities defined by eq. (\ref{watso}).  In this approach
only the estimates of $C^{+}_2$ and $C^{+}_4$ need to be biased with
both the critical temperatures and the exponents, while of course the
estimates of $I^+_{2n+4}$ have to be biased only with the critical
temperatures.  Our final estimates for the amplitudes $C^{+}_{2n} $
are collected in the Table \ref{tab6}. The results of
Ref.[\onlinecite{esshunt}], are reproduced for comparison in Table
\ref{tab7}.

\small
\begin{table}
\caption{ Our final estimates, by first-order DAs, of the critical
amplitudes, $f^+_{\xi}$ of the second-moment correlation length
eq. (\ref{xi2}), and $C^+_{2n}$ of the susceptibilities $\chi_{2n}(K)$
eq. (\ref{2ncorras}), on the HT side of the critical point, for Ising
models with various values $s$ of the spin  on the $sc$ and the $bcc$
lattices and for the lattice scalar field with $\phi^4$
self-interaction. The quartic coupling has the value $g=1.1$ for the
$sc$ lattice, while $g=1.85$ for the $bcc$ lattice. For convenience,
we have reported the value of $C_{2n}/(2n)!$. 
 } 
 \center
\begin{tabular}{|c|c|c|c|c|c|c|c|}
 \hline
 &$s=1/2$& $s=1$ & $s=3/2$ & $s=2$
&$s=5/2$&$s=3$&$\phi^4$\\
 \hline
$bcc$&&&&&&&\\
 \hline
$f^+_{\xi}$& 0.4681(3)&0.4249(1) &0.4107(2)&0.4043(1) &0.4010(1)& 0.3989(2)&
 0.4146(1)\\
$C^+_2/2!$ &0.5202(9)&0.3105(6)&0.2481(5)&0.2186(5)&0.2019(5)&0.1910(4)
&0.2741(7)\\
$C^+_4/4!$&-0.1416(9)&-0.0377(1)&-0.02175(7)&-0.0161(1)&-0.0134(1)
 &-0.01180(8) & -0.02728(8) \\
 $C^+_6/6!$&0.1224(9) & 0.01455(8) & 0.00605(4) &0.00377(6) &0.00282(4)
&0.00231(3) & 0.00862(8) \\
$C^+_{8}/8!$ &-0.150(3) &-0.00798(9)&-0.00240(2)&-0.00125(3) &-0.000845(9)&
 -0.000646(7) & -0.00387(3)  \\
$C^+_{10}/10!$ & 0.22(1) &0.0052(1) &0.00113(1)&0.000497(9)&0.000301(5)&
 0.000215(3) &0.00207(4)\\
$C^+_{12}/12!$ &-0.35(5)&-0.0037(1)&-0.00059(2)&-0.00022(2)
&-0.000119(5)& -0.000079(3) &-0.00123(4) \\
$C^+_{14}/14!$ &0.57(9)&0.0029(2)&0.00032(3)&0.000101(8)&0.000050(3)&
 0.000031(2)&0.00077(5) \\
 \hline
$sc$&&&&&&&\\
 \hline
$f^+_{\xi}$ &0.5070(5) &0.4588(4)&0.4429(4)&0.4356(4) &0.4317(5)
&0.4294(5)&0.4151(1)\\
$C^+_2/2! $ &0.5608(9) &0.338(2) &0.270(2) &0.239(1)&0.220(1)
&0.208(1)&0.2384(7)\\
$C^+_4/4!$&-0.1608(5) &-0.0432(2) &-0.0249(2)&-0.01847(9) &-0.0153(1)
&-0.0135(1)&-0.01595(3)\\
 $C^+_6/6!$&0.146(3) &0.0175(2) &0.00729(4) &0.00454(3)&0.00339(1)
&0.00277(2)&0.00339(1)\\
$C^+_8/8!$ &-0.187(9) &-0.0101(2)&-0.00302(5)&-0.00158(3)&-0.00106(2)
&-0.000809(9)&-0.00102(1)\\
$C^+_{10}/10!$ &0.26(6) &0.0069(3) &0.00148(6) &0.000655(9) &0.000393(9)
&0.000279(9)&0.000367(6)\\
$C^+_{12}/12!$ &-0.19(9)&-0.0049(9)&-0.00079(9)&-0.00030(3)&-0.000162(8)
 &-0.000107(6)&-0.000146(5)\\
$C^+_{14}/14!$ &0.026(9)&0.0016(9)&0.00034(9)&0.00013(4)&0.000066(9)
&0.000041(8)&0.000061(4)\\
 \hline
\colrule  
\end{tabular} 
\label{tab6}
\end{table}

\begin{table}
\caption{ Estimates
 of the amplitudes $C^+_{2n}$ eq. (\ref{2ncorras}),  tabulated in 
 Ref.[\onlinecite{esshunt}] without indication of error and 
only in the case of the Ising model with $s=1/2$.
} 
 \center
\begin{tabular}{|c|c|c|c|c|c|c|}
 \hline
 &$C^+_2/2!$& $C^+_4/4!$& $C^+_6/6!$&$C^+_{8}/8!$
&$C^+_{10}/10!$&$C^+_{12}/12!$\\
 \hline
$sc$ lattice&0.5299 &-0.1530 & 0.1366&-0.1722&0.2601 &-0.4526\\
 \hline 
$bcc$ lattice&0.4952(5)&-0.1385&0.1169 &-0.1397  & 0.2023  & -0.3297 \\ 
 \hline
\colrule  
\end{tabular} 
\label{tab7}
\end{table}

\begin{figure}[tbp]
\begin{center}
\leavevmode
\includegraphics[width=3.37 in]{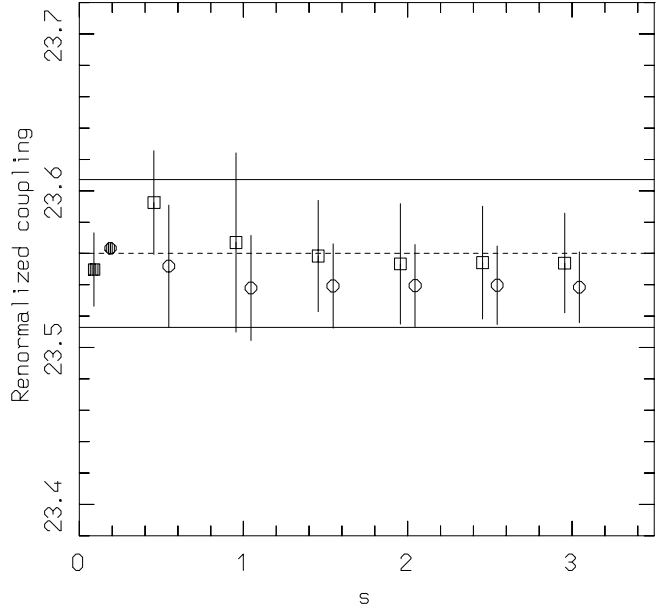}
\caption{\label{rencoup} The renormalized coupling constant $g^+_4$
 for the Ising model with spin $s$ on the $bcc$ lattice (circles) and
 on the $sc$ lattice (squares) vs the value $s$ of the spin.  The HT
 expansions of the Ising systems have been subjected to the variable
 transformation eq.(\ref{rosk}).  For comparison, we have also
 computed $g^+_4$ for the scalar model with the value $g=1.1$ of the
 quartic coupling on the $sc$ lattice (black square) and with $g=1.85$
 on the $bcc$ lattice (black circle). The latter estimates are plotted
 with conventional abscissas near zero.  In all cases, the symbols
 have been slightly shifted apart to avoid superpositions and to keep
 the uncertainty of each estimate visible.
   The
 dashed horizontal line represents the value
 $g^+_4=23.56(3)$ estimated in Ref.[\onlinecite{bcfi4}]. The continuous
 horizontal lines indicate a deviation of $0.2\%$ from the central
 value.
}
\end{center}
\end{figure}

We have estimated the critical values of the HT expansions of the
RCC's either directly, by extrapolation\cite{bcg2n,bcgren} to $K_c^-$
of the simple auxiliary function
\begin{equation} w_{2n}(K)=
(K/K_c)^{\frac{3n-3} {2}}g^+_{2n}(K)
\label{w2n}
\end{equation}
 designed to be regular at $K=0$ and therefore more convenient to
study by DA's, or, more conveniently, but with consistent results,
from the computation of the quantities $r^+_{2n}$ using
eq. (\ref{r2n}).  In Fig. \ref{rencoup}, our estimates for $g^+_4$ are
plotted vs the value $s$ of the spin for Ising systems on the $sc$ and
the $bcc$ lattices and compared to the estimate $g^+_4=23.56(3)$ (dashed line)
 of Ref.[\onlinecite{bcfi4}].  In the same figure,
 we have also shown the values
of $g^+_4$ for the scalar field model on both lattices.
 \begin{table}
\caption{ Our final estimates of the
universal critical values $r^+_{2n}$  of the quantities $r_{2n}(K)$,
with $3 \leq n \leq 7$, for Ising models with various values $s$ of the
spin  and for the scalar field with $\phi^4$
self-interaction, on the $sc$ and the $bcc$ lattices. The quartic
coupling has the value $g=1.1$ for the $sc$
lattice, while $g=1.85$ for the $bcc$ lattice.  We have
also reported the values of the universal amplitude ratios
$I^+_{2n+4}/(2n+2)!$, with $1 \leq n \leq 5$. }
\center
\begin{tabular}{|c|c|c|c|c|c|c|c|}
 \hline
$bcc$ lattice &$s=1/2$& $s=1$ & $s=3/2$ & $s=2$
&$s=5/2$&$s=3$&$\phi^4$\\
 \hline
$g^+_4 $ &23.56(4)&23.54(3) &23.54(3) &23.54(3)&23.54(2)&23.54(2)& 23.56(1)\\
$r^+_6$&2.064(8)&2.061(9)&2.063(5)&2.062(5)&2.062(5)&2.062(5) &2.061(2)\\
 $r^+_8$& 2.54(5) &2.64(5)  &2.61(4) &2.60(4) &2.59(4)&2.58(5) & 2.54(4)\\
$r^+_{10}$ &-15.1(9) &-15.0(5)&-15.4(6)&-15.9(7)&-16.1(8)&-16.3(8) &-15.2(2)\\
$r^+_{12}$ &45(7) &40(5). &44(5)  &48(6) &51(7)  &53(8) &44(3)\\
$r^+_{14}$&1504(240)&1490(115)&1359(82)&1366(100)&1319(100)&1300(100)
&1615(120)\\
$I_6/4!$&0.3307(3)&0.3308(5)&0.3307(2)&0.3307(2)&0.3307(2)&0.3308(2)
&0.3308(1)\\
$I_8/6!$  & 0.2319(6) &0.2319(5)&0.2320(4) &0.2320(4) &0.2320(4)
&0.2320(4) &0.2321(2)\\
$I_{10}/8!$ &0.1667(2) &0.1668(3)&0.1666(2)&0.1664(3)&0.1666(4)&0.1669(5)
&0.1667(2)\\
$I_{12}/10!$ &0.1216(3) &0.1215(1)&0.1216(2) &0.1214(2) &0.1214(3)
&0.1215(2) & 0.1213(3)\\
$I_{14}/12!$ &0.0894(2) &0.0892(2) & 0.0894(3) &0.0894(3)&0.0893(3)
& 0.0893(3) &0.0897(6)\\
 \hline
$sc$ lattice &&&&&&&\\
 \hline
$g^+_4 $ &23.59(4)&23.57(6) &23.56(2)&23.55(4) &23.55(4) &23.55(3)&23.55(3) \\
$r^+_6$& 2.067(11) &2.066(8)&2.064(7) &2.065(7)&2.066(7) &2.065(7)&2.057(3) \\
 $r^+_8$&2.51(7) &2.41(5) &2.45(10)&2.57(10) &2.61(9) &2.61(9) & 2.45(5) \\
$r^+_{10}$ &-17(2) &-14(2)&-14(2)&-14(1)&-14(1) &-14(1) & -15.4(2)\\
$r^+_{12}$ &45(8) & 44(8) &44(6)  &52(4) &54(5)  &51(6) &62(3) \\
$r^+_{14}$ &1460(240)&1390(130)&1644(115) &1477(120)&1362(150)  
&1310(150) &1176(140) \\
$I_6/4!$ &0.3306(5)&0.3306(3) &0.3307(3) &0.3306(3) &0.3306(3) & 0.3306(3)
&0.3310(1)\\
$I_8/6!$ &0.2320(12) &0.2316(7) & 0.2316(8) & 0.2317(8)& 0.2318(8) 
&0.2318(7) &0.2324(3)\\
$I_{10}/8!$ &0.1678(8)&0.1670(10)&0.1665(4)&0.1665(3)&0.1665(3) 
&0.1665(3) &0.1665(5)\\
$I_{12}/10!$&0.1211(6)&0.1214(5) &0.1213(5)&0.1210(5) &0.1209(6) 
&0.1208(4) &0.1206(9)\\
$I_{14}/12!$ &0.0908(12) &0.0906(12) &0.0894(10) &0.0892(10) &0.0889(12) 
&0.0893(8) &0.0884(7) \\
 \hline
\colrule  
\end{tabular} 
\label{tab8}
\end{table}

  Table \ref{tab8} lists our estimates of the quantities $I^+_{2n+4} $
and $r^+_{2n}$, obtained from first- and second-order DA's, for a few
spin-$s$ Ising systems and for the lattice scalar field, on the $sc$
and the $bcc$ lattices. We have imposed that the critical inverse
temperatures take the appropriate values reported in  Table \ref{tab5} and that
an antiferromagnetic singularity is present  at $-K_c$. Only for
the spin-$s$ Ising models, we have taken advantage of the variable
transformation eq.(\ref{rosk}), to reduce the uncertainties of the
estimates and the spread among the central values for different spins. 
We have always
taken care that the uncertainties of our results allow for the errors
of the critical temperatures listed in Table \ref{tab5} and, whenever
the variable transformation eq.(\ref{rosk}) is performed, also for the
error of the exponent $\theta$.

\begin{figure}[tbp]
\begin{center}
\leavevmode
\includegraphics[width=3.37 in]{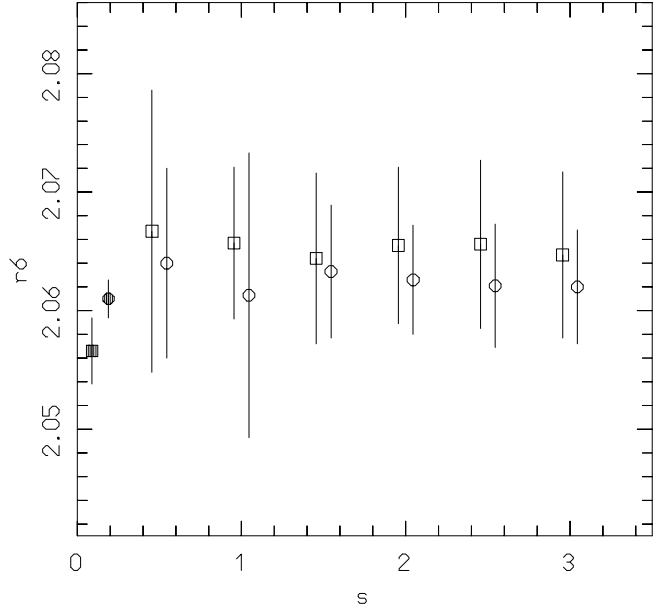}
\caption{\label{r6uni} The ratio $r^+_6$ for the Ising model with
spin $s$ on the $bcc$ lattice (circles), for the Ising model with spin
$s$ on the $sc$ lattice (squares) vs the spin. The expansions for the
Ising systems have been subjected to the variable transformation
eq.(\ref{rosk}). The symbols have been slightly
shifted apart to avoid superpositions and to keep the uncertainties of
each estimate visible. For comparison, we have also computed $r^+_6$ for
the  scalar model with $g=1.1$ on the $sc$ lattice (black
square) and with $g=1.85$ on the $bcc$ lattice (black circle). The
latter estimates are plotted with conventional abscissas near zero. }
\end{center}
\end{figure}

\begin{figure}[tbp]
\begin{center}
\leavevmode
\includegraphics[width=3.37 in]{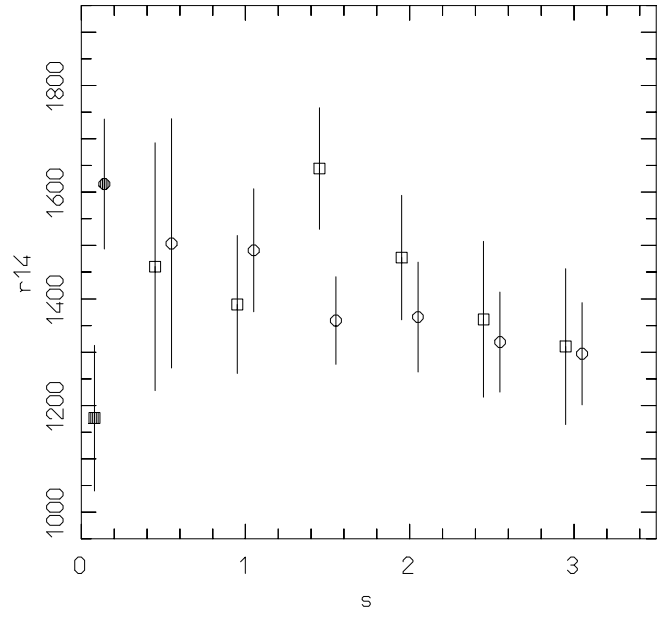}
\caption{ \label{r14uni}  Same as Fig.\ref{r6uni}, but for  $r^+_{14}$.}
\end{center}
\end{figure}

In Figs. \ref{r6uni},...,\ref{r14uni} we have plotted vs the spin, our
estimates of the quantities $r^+_6$, $r^+_8$, $r^+_{10}$, $r^+_{12}$
and $r^+_{14}$ for the Ising models of spin $s=1/2,...3$ on the $sc$
and the $bcc$ lattices. In these figures we have reported, in
correspondence with the conventional value $s=0$ of the abscissa, also
our results for the scalar model in the case of the $sc$ lattice with
quartic coupling $g=1.1$ and, in the case of the $bcc$ lattice, with
coupling $g=1.85$.  The set of estimates shows good universality
properties and moderate relative uncertainties which slowly grow with
$2n$. In the worst case, that of $r^+_{14}$, the uncertainties are
generally not larger than $15\%$.

\begin{table}
\caption{ Our final estimates of the quantities $g^+_4 $, $r^+_6$,
 $r^+_8$, $r^+_{10}$, $r^+_{12}$ and $r^+_{14}$, obtained either from
 the $\phi^4$ results on the $bcc$ lattice or from a weighted average
 of the results on both the $sc$ and the $bcc$ lattices, are compared
 to estimates in the recent literature.  These have been obtained: i)
 from HT expansions\cite{campo2002}, shorter than those analyzed here,
 of the $sc$ lattice scalar field with $\phi^4$ or $\phi^6$
 self-interactions and appropriate couplings; ii) from the
 expansion\cite{gz2} in powers of $\epsilon=d-4$, within the RG
 approach; iii) from the $g-$expansion\cite{gz2,soko} in fixed
 dimension $d=3$, within the renormalization-group approach; iv) from
 various approximations\cite{denjoe,morris,tetra} of the
 renormalization-group equations; v) from MonteCarlo
 simulations\cite{tsyp,kim}.  }
  \center
\begin{tabular}{|c|c|c|c|c|c|c|}
 \hline
& $g^+_4 $&$r^+_6$& $r^+_8$&$r^+_{10}$ &$r^+_{12}$ &$r^+_{14}$\\
 \hline
This work&23.56(1)&2.061(2)&2.54(4) &-15.2(4) &45(5)&1400(200)\\
HT scalar $sc$[\onlinecite{campo2002}] &23.56(2)& 2.056(5)&2.3(1)&-13(4)&&\\
$\epsilon$-exp.[\onlinecite{gz2}]&23.3&2.12(12) &2.42(30)&-12(1)&&\\
$g$-exp.[\onlinecite{gz2}]&23.64(7)&2.053(8)&2.47(25)&-25(18)&&\\
$g$-exp.[\onlinecite{soko}]&23.71&2.060&&&&\\
Approx. RG[\onlinecite{denjoe}]&&1.938&2.505&-12.599&10.902&\\
Approx. RG[\onlinecite{morris}]&20.72(1)&2.063(5)&2.47(5)&-19(1)&&\\
Approx. RG[\onlinecite{tetra}]&28.9&1.92&2.17&&&\\
MC  Ising $sc$[\onlinecite{tsyp}] &23.3(5)&2.72(31)&&&&\\
MC  Ising $sc$[\onlinecite{kim}] &24.5(2)&3.24(24)&&&&\\
 \hline
\colrule  
\end{tabular} 
\label{tab9}
\end{table}

In Table \ref{tab9}, we have collected our final estimates of the
ratios $r^+_6$, $r^+_8$,... $r^+_{14}$, obtained either by simply choosing
our result for the scalar field system on the $bcc$ lattice, as in the case
of the lowest-order ratios, or from a weighted average of the
estimates on the $sc$ and $bcc$ lattices for the same system, as in the
case of the largest-order ratios.  Our values are compared with the
estimates already obtained in the recent literature by various
methods, including the analysis of significantly shorter HT expansions.
\section{ Conclusions}
For a wide class of models in the 3D Ising universality class, we have
described properties of the higher susceptibilities on the HT side of
the critical point, which are relevant for the construction of
approximate representations of the critical ES.  We have based on
high-temperature and low-field bivariate expansions, that we have
significantly extended or computed ``ex novo''.  The models under
scrutiny include the conventional Ising system with spin $s=1/2$, the
Ising model with spin $s>1/2$ and the lattice scalar field, defined on
the three-dimensional $sc$ and $bcc$ lattices.  In this paper our HT
data have been used to improve the accuracy and confirm the overall
consistency of the current description of these models in critical
conditions, by testing simple predictions of the scaling hypothesis as
well as the validity of the universality property of the gap exponent
and of appropriate combinations of critical amplitudes.  Some of these
tests are presently feasible only within a series approach.  Our main
result 
is a set of more accurate estimates of the first three already known
$r^+_{2n}$ parameters and a computation of two additional ratios, which
enable us to formulate an update of the parametric form of the ES.

At the order of expansion reached in our study, we still observe a
small residual spread of the estimates of the gap exponent and of the
ratios $r^+_{2n}$, around the predictions of asymptotic scaling and
universality.  This fact is readily explained  by the obvious
limitations of our numerical analysis: namely  the still relatively 
moderate span of our expansions, in spite of their significant extension,
 the notoriously slower convergence of
the expansions in the case of the $sc$ lattice and the incomplete
allowance of the non-analytic corrections to scaling by the current
tools of series analysis. 
\section{Acknowledgements}
We are deeply grateful to Riccardo Guida for his precious advise and
for his generous help in the revision of our draft.  We have enjoyed the
hospitality and support of the Physics Depts. of Milano-Bicocca
University and of Milano University.  Partial support by the MIUR is
acknowledged.

\footnotesize

\newpage

\begin{figure}[tbp]
\begin{center}
\leavevmode
\includegraphics[width=3.37 in]{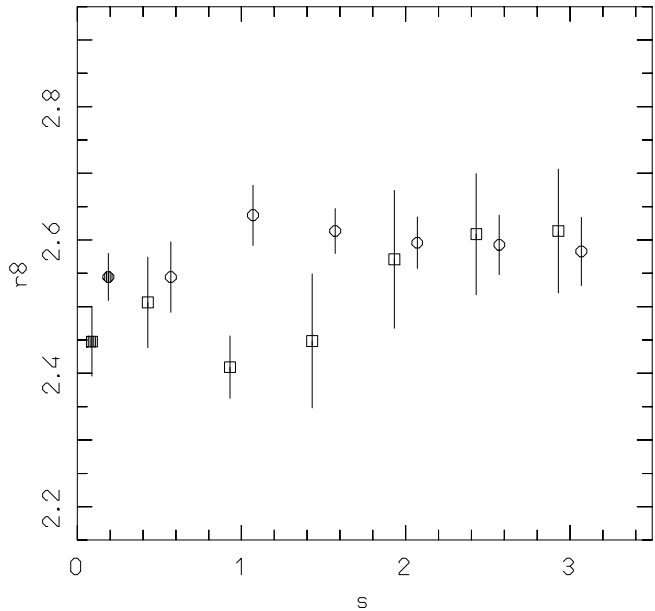}
\caption{\label{r8uni}  Same as Fig.\ref{r6uni}, but for  $r^+_8$.}
\end{center}
\end{figure}

\begin{figure}[tbp]
\begin{center}
\leavevmode
\includegraphics[width=3.37 in]{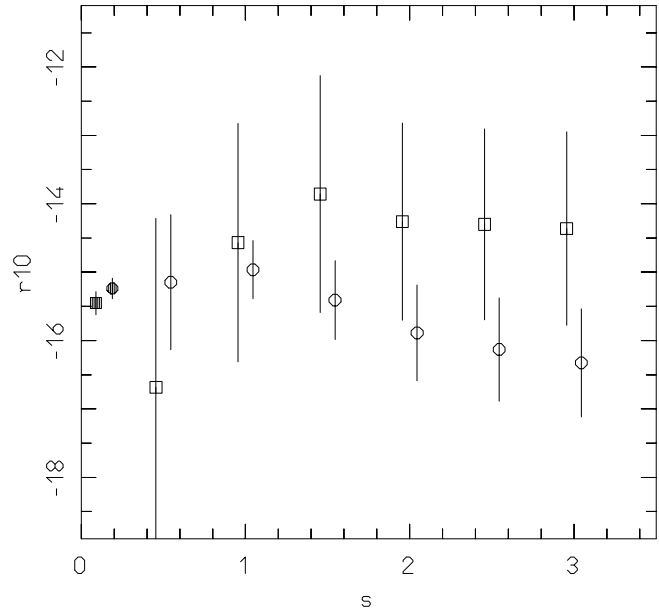}
\caption{\label{r10uni}  Same as Fig.\ref{r6uni}, but for  $r^+_{10}$.}
\end{center}
\end{figure}

\begin{figure}[tbp]
\begin{center}
\leavevmode
\includegraphics[width=3.37 in]{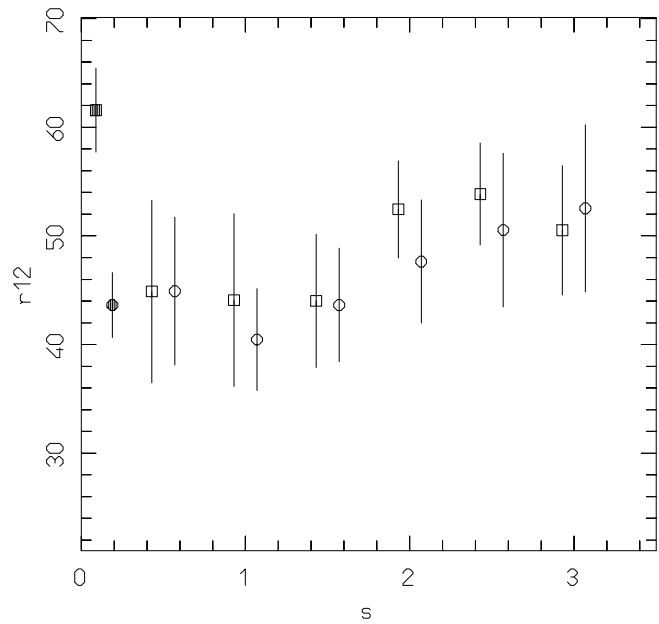}
\caption{ \label{r12uni}  Same as Fig.\ref{r6uni}, but for  $r^+_{12}$.}
\end{center}
\end{figure}


\begin{thebibliography}{}
\bibitem{essam} Shorter expansions had been previously studied in 
Ref.[\onlinecite{esshunt}], but the series coefficients were not published.
\end{thebibliography}
\end{document}